\documentclass[11pt, a4paper, onecolumn, copyright]{AweAI}

\usepackage{multirow}
\usepackage[authoryear, sort&compress, round]{natbib}
\usepackage{float}
\usepackage{graphicx}
\usepackage{xspace} 
\definecolor{lightblue}{HTML}{EBF3FF} 
\usepackage[table]{xcolor} 
\usepackage{tcolorbox}
\usepackage{listings} 

\tcbuselibrary{listings, skins, breakable}

\newtcblisting{markdownbox}[1][]{
  listing engine=listings,
  listing options={
    basicstyle=\small\ttfamily,
    breaklines=true,
    breakatwhitespace=true,
    numbers=none,
    frame=none, 
    columns=fullflexible,
    showstringspaces=false,
    xleftmargin=0pt,
    xrightmargin=0pt,
    aboveskip=0pt,
    belowskip=0pt
  },
  colback=gray!5!white,
  colframe=gray!70!black,
  title={\textbf{Prompt (Markdown)}},
  fonttitle=\bfseries,
  listing only,
  left=5mm, right=5mm, top=3mm, bottom=3mm,
  breakable,
  enhanced,
  overlay={\begin{tcbclipinterior}\fill[gray!5!white] (frame.south west) rectangle ([xshift=0mm]frame.north west);\end{tcbclipinterior}},
  #1
}
\definecolor{bg}{rgb}{0.95,0.95,0.95}
\uselogo{}

\DeclareRobustCommand{\github}{\raisebox{-1.5pt}{\includegraphics[height=1.05em]{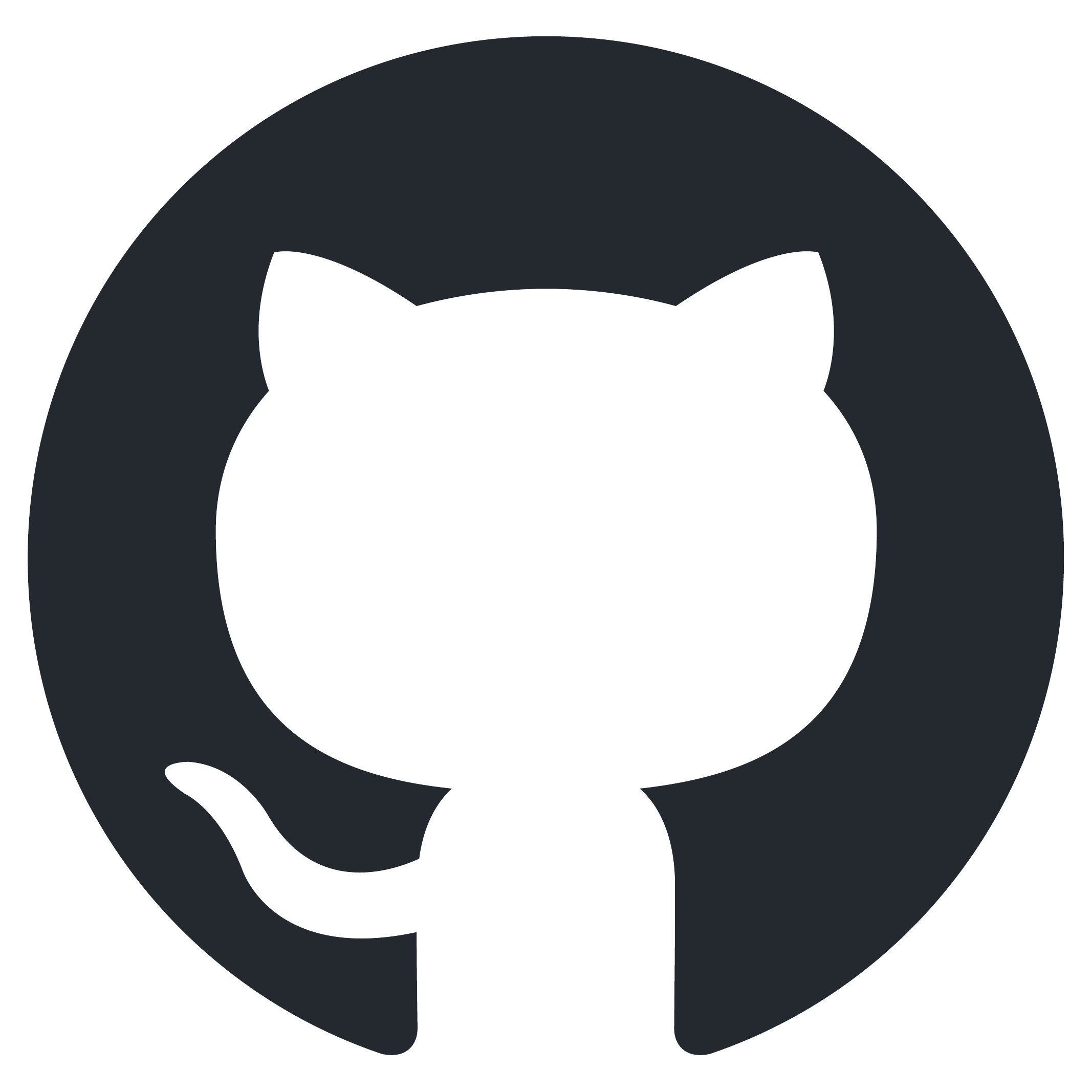}}\xspace}

\DeclareRobustCommand{\hfdataset}{\raisebox{-1.5pt}{\includegraphics[height=1.05em]{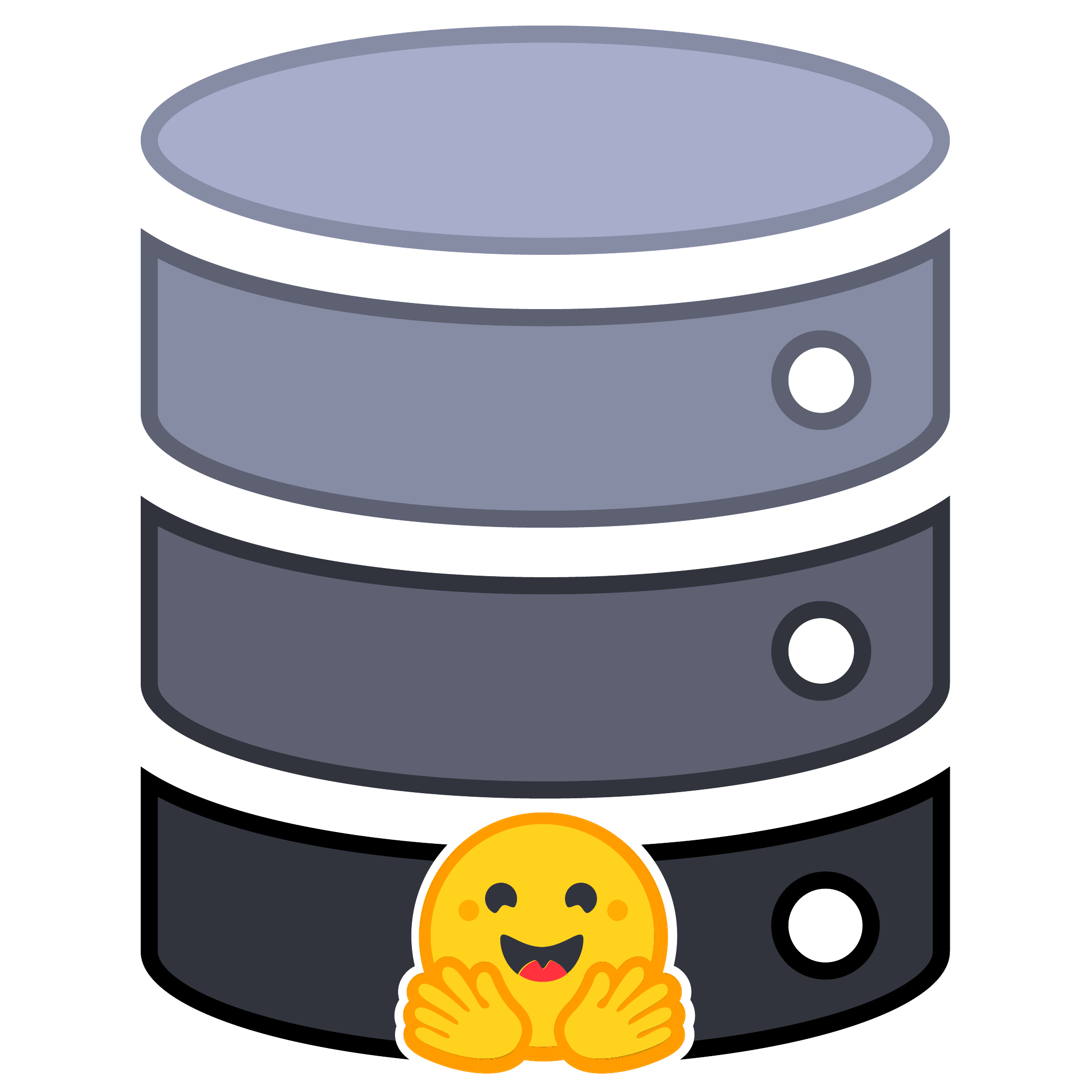}}\xspace}

\newcommand{\footerlinks}{%
  \hfdataset\ \href{https://huggingface.co/collections/AweAI-Team/denovoswe}{Dataset}\hspace{1.5em}%
  \github\ \href{https://github.com/AweAI-Team/DeNovoSWE}{GitHub}%
}

\title{
DeNovoSWE: Scaling Long-Horizon Environments for Generating Entire Repositories from Scratch
}

\newcommand{\publicday}{Jun.~10, 2026}

\author[*]{Jiale Zhao}
\author[*]{Guoxin Chen}
\author[ \hspace{-0.3em}]{Fanzhe Meng}

\author[ \hspace{-0.3em}$^\dag$]{Wayne Xin Zhao}
\author[ \hspace{-0.3em}$^\dag$]{Ruihua Song}
\author[ \hspace{-0.3em}]{Ji-Rong Wen}
\author[ \hspace{-0.3em}$^\dag$]{Kai Jia}

\correspondingauthor{\{marshmallowzjl, gx.chen.chn, batmanfly\}@gmail.com, songruihua\_bloon@outlook.com, jiakai@bytedance.com}

\affil[1]{Gaoling School of Artificial Intelligence, Renmin University of China}
\affil[2]{Independent Researcher}
\affil[3]{AweAI Team\footnote{$^*$Equal Contributions. $^\dag$Corresponding authors.\hfill\textbf{Date:} \publicday.}}

\begin{abstract}
As the capabilities of LLM-based code agents continue to advance, their expected role is expanding beyond localized bug fixing in existing codebases toward architecting and implementing complete software repositories from high-level specifications. However, training agents for such long-horizon software engineering tasks remains difficult due to the scarcity of large-scale, verifiable whole-repository generation data.
In this paper, we introduce \textbf{DeNovoSWE}, a large-scale dataset for whole-repository generation. DeNovoSWE comprises 4,818 high-quality instances, where each instance requires generating a complete repository from documentation.
Our dataset is automatically constructed through a carefully designed sandboxed agentic workflow, enabling scalable curation without human annotation. DeNovoSWE is constructed with "divide and conquer" and critic-repair philosophy. 
To balance data quality and diversity, we further introduce a difficulty-aware trajectory filtering strategy.
Fine-tuning Qwen3-30B-A3B on DeNovoSWE substantially improves long-horizon SWE performance, raising its score on the challenging BeyondSWE-Doc2Repo benchmark from 5.8\% to 47.2\%.

\centerline{\footerlinks}
\end{abstract}

\makeatletter
\begin{document}

\begingroup
\makeatletter
\renewcommand{\thefootnote}{}
\renewcommand{\@makefnmark}{}
\maketitle
\makeatother
\endgroup

\begin{figure*}[h] 
  \begin{center}
    \centerline{\includegraphics[width=1\textwidth]{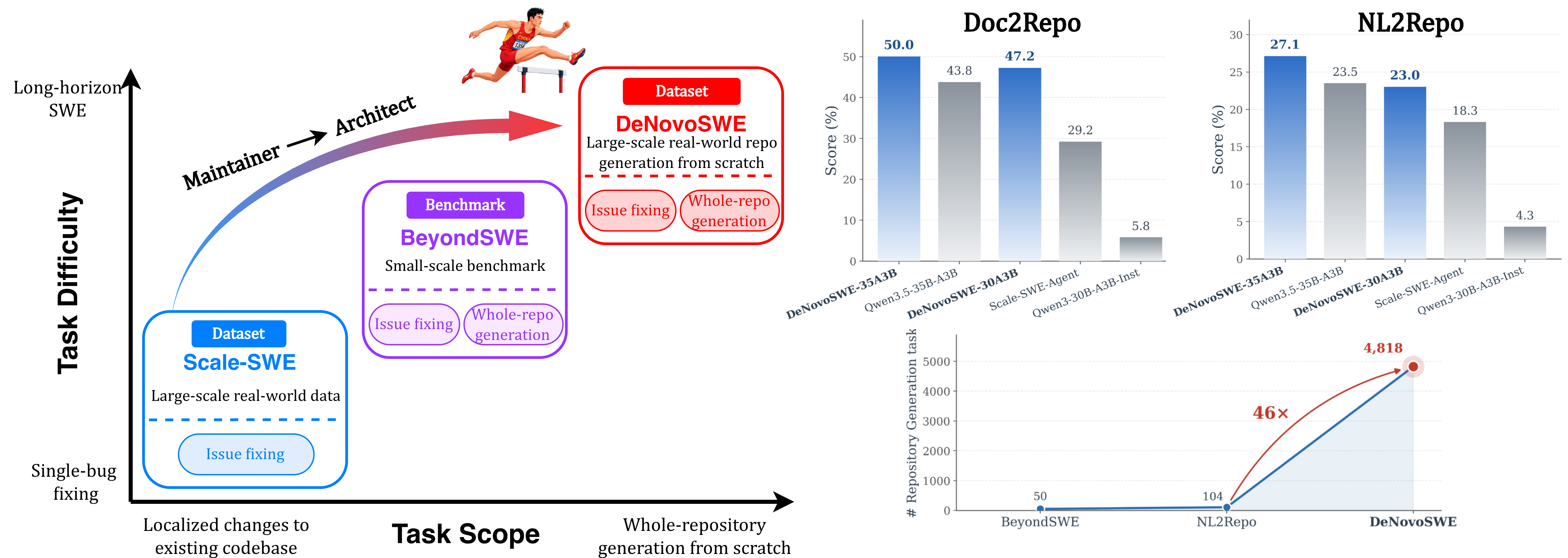}} 
    \caption{
Overview of DeNovoSWE and its role in scaling long-horizon software engineering tasks.
Left: DeNovoSWE extends prior SWE datasets along both task scope and task difficulty, moving from localized issue fixing in existing codebases to whole-repository generation from scratch, thereby requiring agents to shift from maintainer-like localized editing toward architect-level repository construction.
Right: DeNovoSWE provides substantially larger-scale repository-generation supervision, containing 4,818 tasks, about 46$\times$ larger than NL2Repo.
    }
    \label{system}
  \end{center}
  \vspace{-0.6cm} 
\end{figure*}

\section{Introduction}


LLM-based code agents have garnered significant attention for their demonstrated potential in tackling complex software engineering (SWE) tasks~\citep{anthropic2025claudesonnet45, googledeepmind2026gemini3pro,openai2026gpt54}, as reflected by their strong performance on benchmarks such as SWE-bench~\citep{jimenez2023swe}. However, as frontier models achieve increasingly high scores on SWE-bench-Verified, the benchmark now faces a dual limitation: it is becoming less discriminative among strong code agents, and its predominantly issue-level tasks do not sufficiently stress long-horizon repository-level reasoning and implementation~\citep{openai2026swebenchverified}. Long-horizon capabilities are crucial for solving complex real-world tasks~\citep{chen2025iterresearch,tang2026agent,chen2026toward}. Furthermore, recent benchmarks~\citep{chen2026beyondswe,ding2025nl2repo,deng2025swe} have emerged to systematically evaluate long-horizon coding ability. These benchmarks emphasize long-horizon coding ability and better reflect real-world SWE workflows. Despite this progress, development in long-horizon coding ability remains constrained by the scarcity of training data, since existing SWE training datasets are still largely centered on single-issue bug fixing.

Recently, there have been several works that automatically scale the construction of real-world SWE training data~\citep{zhao2026immersion,badertdinov2025swe,tao2026swe,fu2026davinci}.
Yet, most existing SWE training datasets are still largely centered on to single-issue fixes rather than whole-repository generation — a far more demanding setting that requires long-horizon planning and complex, interdependent coding. When tasked with generating entire repositories, even state-of-the-art agents fall short, as shown by recent benchmarks like BeyondSWE and NL2RepoBench~\citep{chen2026beyondswe,ding2025nl2repo}. This is primarily limited by the lack of long-horizon verifiable SWE data. 
Despite the success of these works on scaling up SWE data, they are mostly limited to issue resolution, which is insufficient for training agents on long-horizon repository-level tasks. 

Scaling whole-repository generation as a verifiable SWE task requires addressing three core challenges: documentation construction, evaluation design, and leakage-free task execution. Although GitHub hosts a vast number of real-world repositories, their existing documentation is often incomplete, unstructured, or misaligned with executable behavior, making it insufficient for directly defining document-to-repository generation tasks. Constructing documentation that is well organized, sufficiently comprehensive, and consistent with the behavior expected by the evaluation suite is therefore a central challenge, as it directly determines task validity and data quality. Beyond documentation, whole-repository generation also requires an objective and scalable evaluation protocol that can assess diverse repositories through executable tests. Finally, because modern code agents operate in interactive environments with tool access, they may exploit unintended leakage channels to access the reference implementation. Robust sandboxing and containment policies are therefore necessary to prevent oracle leakage during execution.

To address these challenges, we present \textbf{DeNovoSWE}, a large-scale dataset for long-horizon software engineering that requires agents to generate complete repositories from documentation. We introduce an automated sandboxed pipeline for constructing high-quality document-to-repository task instances. To synthesize documentation that is comprehensive, well organized, and consistent with the target repository behavior and executable evaluation suite, our framework adopts a divide-and-conquer paradigm coupled with an iterative critic-repair mechanism. To further balance trajectory quality and task diversity, especially for complex instances where fully successful trajectories are difficult to obtain, we propose a difficulty-aware trajectory filtering strategy for curating verifiable training trajectories. Comprehensive evaluations on BeyondSWE-Doc2Repo~\citep{chen2026beyondswe} and NL2RepoBench~\citep{ding2025nl2repo} show that training on DeNovoSWE substantially improves model performance on whole-repository generation tasks. These results suggest that DeNovoSWE fills a critical gap in verifiable long-horizon SWE training data for repository-scale generation.

To summarize, our main contributions are as follows:

\begin{itemize}[leftmargin=*, noitemsep, topsep=0pt]
\item We introduce an automated sandboxed pipeline for constructing document-to-repository data at scale, resulting in \textbf{DeNovoSWE-Data}, a large-scale whole-repository generation dataset for long-horizon software engineering, comprising 4,818 high-quality instances.
\item We propose a difficulty-aware trajectory filtering mechanism that balances the trade-off between data quality and task diversity, enabling effective curation of verifiable expert trajectories for complex repository-generation tasks.
\item We develop \textbf{DeNovoSWE-Agent} by training on DeNovoSWE-Data, and empirically show that it substantially improves model performance on whole-repository generation from documentation.
\end{itemize}

\section{Related Work}
\textbf{SWE Benchmark.} Since the introduction of the prevailing software engineering benchmark, SWE-bench~\citep{jimenez2023swe} and SWE-bench-Verified~\citep{chowdhury2024introducing}, many other benchmarks have emerged to assess multi-modal~\citep{yang2024swe}, multi-language~\citep{zan2025multi,rashid2025swe,guo2025omnigirl}. and long-horizon capabilities~\citep{deng2025swe}. 
Because of the importance of long-horizon SWE tasks, there have emerged some benchmarks that evaluate code agent performance in generating the whole repository from scratch, like BeyondSWE~\citep{chen2026beyondswe}, NL2Repo~\citep{ding2025nl2repo}, and ProgramBench~\citep{yang2026programbench}.

\textbf{SWE Datasets.} High-quality data is pivotal for enhancing the programming capabilities of Large Language Models (LLMs). Recently, there has been a surge in repository-level software engineering datasets aimed at addressing complex coding tasks. SWE-Gym~\citep{pan2024training} focuses on constructing real-world SWE data. Many large repository level data has then been proposed, like Scale-SWE~\citep{zhao2026immersion}, OpenSWE~\citep{fu2026davinci}, and SWE-rebench~\citep{badertdinov2025swe}.

\textbf{SWE Models.} Recent advancements have introduced powerful models specialized for SWE tasks, including SWE-RL~\citep{wei2025swe}, SWE-Swiss~\citep{SWESwiss2025}, SWE-World~\citep{sun2026swe}, and SWE-Master~\citep{song2026swe}, Orchard~\citep{peng2026orchardopensourceagenticmodeling}.
In parallel, frameworks such as SWE-agent~\citep{yang2024sweagent}, Mini-SWE-Agent~\citep{yang2024sweagent}, OpenHands~\citep{wang2025openhands}, OpenComputer~\citep{wei2026opencomputerverifiablesoftwareworlds}, and MOpenHands~\citep{zan2025multiswebench} serve as effective scaffolds to streamline interactions with development environments.

\section{DeNovoSWE: Scaling Long-Horizon Repository Generation}

DeNovoSWE is built on a sandboxed multi-agent system that follows a structured workflow for constructing verifiable document-to-repository tasks. The central challenge is to generate documentation that is comprehensive, well organized, and executable in the sense that it provides sufficient behavioral and structural information for an implementation generated from it to pass the repository's evaluation tests. This requirement is substantially more demanding than ordinary documentation synthesis, as the documentation must faithfully capture repository-level functionality, interfaces, dependencies, and interactions across components. To address this challenge, we adopt a \textbf{divide-and-conquer} methodology: each repository is decomposed into distinct capabilities, and sandboxed agents generate targeted documentation for each capability in a modular manner. Since producing complete and test-aligned documentation in a single pass is difficult, we further introduce an iterative \textbf{critic-repair} mechanism that identifies omissions and inconsistencies and revises the documentation accordingly. Unless otherwise specified, all agent modules and LLM-as-judge components are implemented with GPT-5.4 and GPT-5.5.



\begin{figure*}[t!] 
  \begin{center}
    \centerline{\includegraphics[width=1\textwidth]{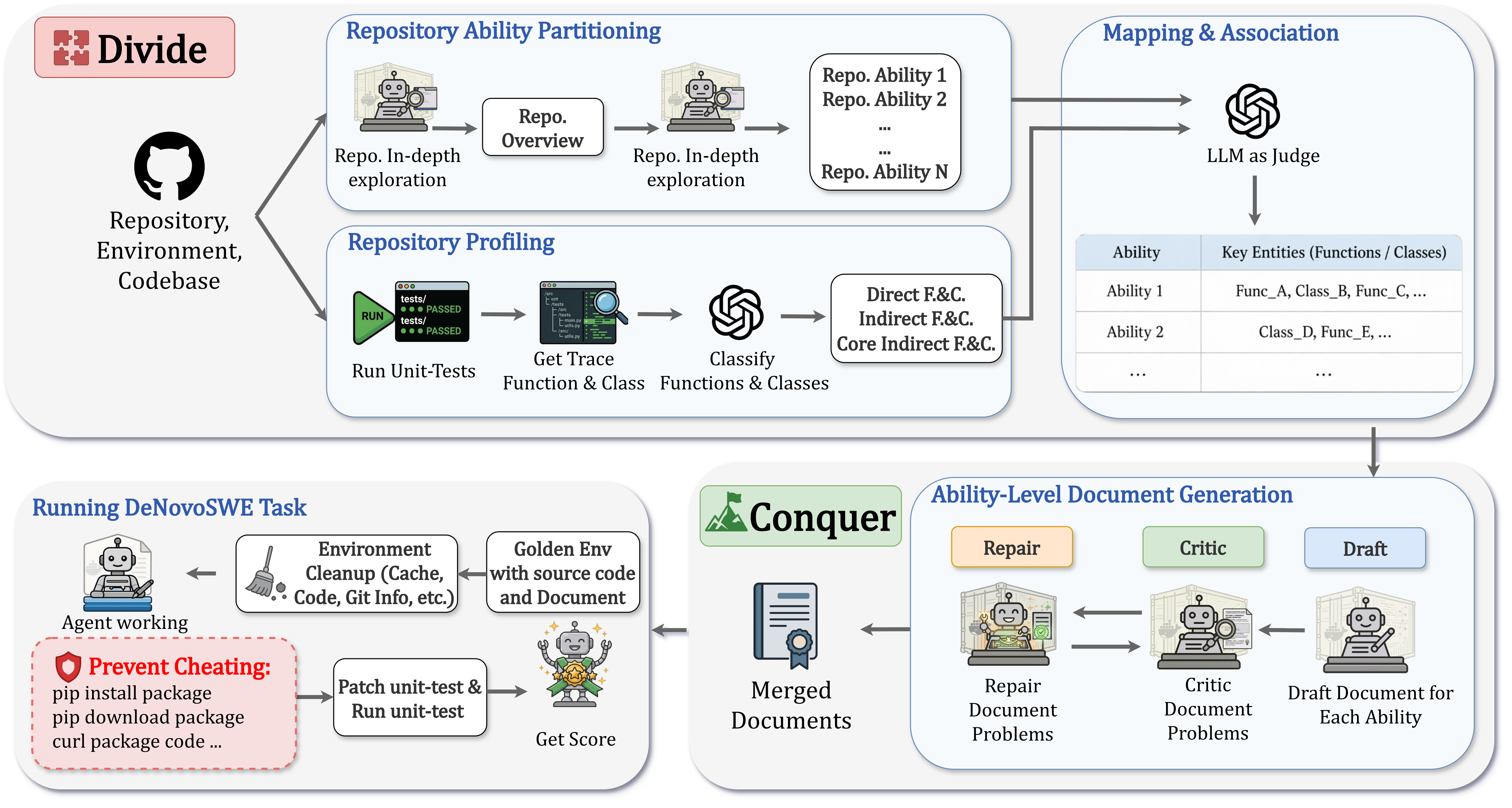}} 
    \caption{
Overview of the DeNovoSWE framework based on a Divide-and-Conquer design. In the Divide Phase (Top), the repository is decoupled via concurrent tracks: Repository Ability Partitioning for capability extraction and Repository Profiling for code-dependency tracing. These aspects are consolidated through an LLM-as-a-Judge to map high-level abilities onto specific code structures. F. \& C. denotes functions and classes. In the Conquer Phase (Bottom), an iterative multi-agent pipeline (Draft-Critic-Repair) is executed for Ability-Level Document Generation. The resulting merged documentation is fed into a sandboxed Golden Environment, where the software agent's performance is rigorously benchmarked under strict network and package deployment constraints to determine the final evaluation score.
    }
    \label{system}
  \end{center}
  \vspace{-0.6cm} 
\end{figure*}
\subsection{Divide}\label{sec:divide}

The objective of this phase is to decompose a repository into distinct functional capabilities and associate each capability with its relevant modules, functions, classes, and interfaces. This decomposition provides downstream documentation agents with clearer and more localized requirements, thereby reducing the complexity of repository-level documentation generation.

\textbf{Repository capability decomposition.} We first employ an overview writer agent to explore the entire repository and generate a high-level summary of its overall purpose, architecture, and major components. This overview serves both as the introduction to the final documentation and as global context for subsequent capability-level decomposition. Next, a capability writer agent identifies the major functional capabilities of the repository and assigns relevant implementation units, such as modules, functions, classes, and public interfaces, to each capability. Each identified capability is then organized as a dedicated chapter in the final documentation. This modular decomposition improves coverage of repository functionality while imposing a structured organization on the generated document.

\textbf{Repository profiling.} This step identifies the functions and classes that should be prioritized in the generated documentation. We first execute the repository's unit tests and capture runtime traces to identify implementation units exercised by the test suite. We then categorize these functions and classes into three groups: direct, core indirect, and non-core indirect components. Direct components are those explicitly imported, instantiated, or invoked in the unit test files. These components are documented with sufficient detail, especially their import paths, public APIs, input-output behavior, and expected usage, since omitting them can make the document-to-repository task under-specified. Indirect components are not directly referenced by the tests but are reached through the execution traces of direct components. Among them, core indirect components are those that affect observable behavior or are necessary for reproducing the tested functionality, and are therefore included in the documentation. In contrast, non-core indirect components correspond to internal implementation details that are not essential for the evaluated behavior. We exclude these components from the documentation, leaving their implementation flexible and thereby preserving both task diversity and realistic repository-level challenge.

\textbf{Mapping \& association.} This stage associates the profiled functions and classes with their corresponding repository capabilities. The resulting mappings are then provided to the capability documentation writer agent, enabling it to generate targeted documentation for each capability. Specifically, we use an LLM-based classifier to assign all components to capabilities. To improve classification accuracy, the classifier is provided with rich contextual information, including the source code and file path of each component, the repository file structure, the complete list of identified capabilities, and the high-level repository overview. This mapping step ensures that capability-level documentation is grounded in the relevant implementation units rather than generated from coarse repository summaries alone.

\subsection{Conquer}\label{sec:conquer}

This stage consists of three sandboxed agents: a draft agent, a critic agent, and a repair agent. The agents process one repository capability at a time and iteratively refine the corresponding capability-level documentation. After all capabilities have been processed, the resulting capability-level documents are merged with the repository overview to form the final task documentation.

\textbf{Draft agent.} The draft agent generates an initial document for each repository capability. During dataset construction, the agent operates in a sandboxed environment with access to the source repository, allowing it to inspect relevant code when additional context is needed. For a target capability $a_i$, the initial draft document $D_i^{(0)}$ is generated as:
$$D_i^{(0)} = f_{\mathrm{draft}}(a_i, \mathcal{A}, O, M_i, \mathcal{E})$$
where $\mathrm{A}$ denotes the full set of identified repository capabilities, $O$ is the repository overview, $M_i$ contains the functions, classes, and interfaces mapped to capability $a_i$, and $\mathcal{E}$ denotes the sandboxed execution environment.

\textbf{Critic agent.} The critic agent identifies deficiencies in either the initial draft or the intermediate versions revised by the repair agent. For each capability-level document, it evaluates the structural organization, checks whether the relevant direct and core indirect components are sufficiently covered, and detects missing or under-specified APIs, import paths, input-output behavior, and usage constraints. Importantly, the critic agent assesses whether the current document provides enough information for a downstream implementation agent to reproduce the evaluated functionality, while avoiding excessive implementation details that would make the task trivial or leak test-specific behavior.

Formally, let $D_i^{(t)}$ denote the documentation for capability $a_i$ at iteration $t$, where $D_i^{(0)}$ is the initial draft produced by the draft agent. The criticism $C_i^{(t)}$ is generated as:
$$C_i^{(t)} = f_{\mathrm{critic}}(D_i^{(t)}, M_i^{\mathrm{miss}}, O, \mathcal{A}, a_i, \mathcal{E})$$
where $M_i^{\mathrm{miss}}$ denotes the set of direct and core indirect components that are missing or insufficiently described in $D_i^{(t)}$, as detected by a rule-based coverage verification procedure. Here, $\mathcal{E}$ is the sandboxed execution environment, $O$ is the repository overview, and $\mathcal{A}$ is the full set of identified repository capabilities, following the definitions above. The resulting criticism $C_i^{(t)}$ is then passed to the repair agent for targeted refinement.

\textbf{Repair agent.} The repair agent addresses the deficiencies identified by the critic agent. Similar to the draft agent, it operates in a sandboxed environment during dataset construction, allowing it to inspect the source repository and retrieve the technical context needed for targeted revision.

Formally, at iteration $t$, the repair agent takes as input the current document version $D_i^{(t)}$, the criticism $C_i^{(t)}$, the target capability $a_i$, the full capability set $\mathcal{A}$, the documentation-relevant components $M_i$, the missing or under-specified components $M_i^{\mathrm{miss}}$, the repository overview $O$, and the sandboxed environment $\mathcal{E}$. The repaired document is generated as:
$$D_i^{(t+1)} = f_{\mathrm{repair}}(D_i^{(t)}, C_i^{(t)}, a_i, \mathcal{A}, M_i, M_i^{\mathrm{miss}}, O, \mathcal{E})$$
The output $D_i^{(t+1)}$ serves as the updated capability-level documentation. If the critic still identifies substantial omissions or inconsistencies and the iteration budget has not been exhausted, the document is passed back into the critic-repair loop; otherwise, it is finalized and later merged into the complete repository documentation.

\subsection{Evaluation Protocol and Leakage Prevention}\label{sec:eval}
Each complete DeNovoSWE instance comprises the core elements detailed in Table~\ref{tab:data_structure}. 

\textbf{Pre-cleanup.}
In the default Docker image, the codebase of the target repository is initially fully intact to grant users maximum flexibility when deploying DeNovoSWE. However, to evaluate document-to-repository generation, a rigorous cleanup process must be executed to establish a clean environment. This pre-cleanup phase consists of the following operations:
\begin{itemize}[leftmargin=*, noitemsep, topsep=0pt]
    \item \textbf{Environment Preservation and Source Stripping:} This process precisely identifies and retains the static build configurations and runtime dependencies of the original repository while entirely stripping out the existing code implementations and test suites.
    \item \textbf{Multi-channel Leak Purging:} To prevent the LLM agent from cheating, the cleanup script thoroughly sweeps across multiple Python environments to erase \texttt{site-packages} traces, hidden \texttt{pip} wheel caches, and transient compilation artifacts in \texttt{/tmp} (such as intermediate C/Rust extension wrappers), thereby severing all leakage channels for source code recovery.
    \item \textbf{Git history sanitization:} By completely destroying and re-initializing the \texttt{.git} directory, this step eliminates the vulnerability where agents could exploit \texttt{git reflog} or loose Git objects to reverse-engineer and reconstruct the original repository history, ensuring a true ``closed-book'' generation task.
\end{itemize}

\textbf{Runtime cheating prevention.} Although the pre-cleanup phase removes Git history, local caches, pre-installed package copies, and other residual artifacts, thereby preventing direct access to the reference implementation within the local environment, an agent with shell access may still attempt to recover the original source through external channels. For example, it may try to clone the reference repository from public hosting services, install or download the target package via \texttt{pip install} or \texttt{pip download}, or retrieve source files through network utilities such as \texttt{curl} or \texttt{wget}. To mitigate these risks, we enforce a command restriction policy inside the Docker container that blocks common source-recovery operations and network-based retrieval paths targeting the reference repository or package. In addition, we apply a rule-based fuzzy-matching filter using the \texttt{pypi\_name} field in DeNovoSWE to detect suspicious commands that reference the target package or related source-distribution artifacts. Beyond these automated checks, we further audit agent execution traces with an LLM-as-judge procedure to identify potential circumvention attempts that are difficult to capture with handcrafted rules alone.

\subsection{Repository Selection.}

We construct executable Docker environments for candidate repositories using the \textbf{Scale-SWE} framework~\citep{zhao2026immersion}. Following its repository-level filtering and environment-building protocol, we further expand the candidate pool with additional real-world repositories. To ensure environment stability and reliable evaluation, we first run the original unit-test suite in each constructed environment and filter out repositories whose test pass rate is below 90\%. We then measure the test coverage over the original source code and retain only repositories with coverage above 50\%. This selection process ensures that the resulting repositories are both executable and sufficiently constrained by behavioral tests, making them suitable for constructing verifiable document-to-repository generation tasks.

\subsection{Dataset Statistics and Analysis}\label{sec:stats}
\begin{table}[h]
    \centering
    \caption{Detailed statistics of DeNovoSWE data summary. We report the mean and percentiles (P50, P75, P90) together with the maximum values for each metric.}
    \label{tab:patch_extraction_stats}
    \vspace{5pt}
    \begin{tabular}{lrrrrr}
        \toprule
        \textbf{Metric} & \textbf{Mean} & \textbf{P50} & \textbf{P75} & \textbf{P90} & \textbf{Max} \\
        \midrule
        Unit-test Count & 205.0 & 79.0 & 197.0 & 464.0 & 8903 \\
        Test Files Count & 31.5 & 12.0 & 27.0 & 59.3 & 8807 \\
        Coverage Percent & 85.5 & 89.6 & 96.5 & 99.9 & 100 \\
        Measured Files & 19.3 & 9.0 & 21.0 & 42.0 & 1995 \\
        \bottomrule
    \end{tabular}
\end{table}

To provide a comprehensive understanding of DeNovoSWE, we analyze its scale and fine-grained characteristics, and further compare it with existing repository-generation datasets and benchmarks.

\textbf{Comparison with existing benchmarks.} As shown in Figure~\ref{system}, DeNovoSWE substantially expands the scale of repository-generation tasks. Prior benchmarks such as NL2RepoBench~\citep{ding2025nl2repo} and BeyondSWE-Doc2Repo~\citep{chen2026beyondswe} contain 104 and 50 task instances, respectively, limiting the diversity. In contrast, DeNovoSWE contains 4,818 instances, making it over an order of magnitude larger than existing repository-generation benchmarks.

\textbf{Fine-grained metric distribution.} Table~\ref{tab:patch_extraction_stats} reports fine-grained statistics of DeNovoSWE instances, including test coverage, measured source files, unit-test count, and test-file count. DeNovoSWE exhibits strong executable coverage, with an average coverage of 85.5\% and a median of 89.6\%; the coverage further increases to 96.5\% at P75 and 99.9\% at P90. This indicates that the selected repositories are generally well constrained by unit tests, providing a reliable basis for verifiable document-to-repository generation. Beyond test quality, DeNovoSWE also captures substantial repository-level complexity. The unit tests exercise a median of 9 source files, increasing to 21 at P75 and 42 at P90, suggesting that the evaluated behavior often spans multiple files and interacting components rather than isolated functions. The dataset also contains a median of 79 unit tests distributed across 12 test files, further showing that evaluation is based on diverse executable checks. Together, these statistics demonstrate that DeNovoSWE instances are well suited for studying long-horizon repository generation under broad behavioral constraints.

\section{Difficulty-Aware Trajectory Filtering}
For conventional issue-level SWE tasks, data construction typically retains only trajectories that pass all unit tests. However, this criterion becomes overly restrictive for document-to-repository generation. Because the task requires reconstructing an entire repository, even strong agents often suffer from compounding errors across files, APIs, dependencies, and implementation details, making fully successful trajectories difficult to obtain. As a result, generated trajectories exhibit varying degrees of correctness, which we quantify by the unit-test pass ratio:

$$score = \frac{N_{\text{passed}}}{N_{\text{total}}}$$

where $N_{\text{passed}}$ represents the number of passed unit tests and $N_{\text{total}}$ represents the total number of unit tests.

This raises a key question: how should high-quality trajectories be selected when their scores vary continuously between 0 and 1? A straightforward solution is to apply a fixed score threshold, such as $0.95$. However, such static filtering conflates trajectory quality with task difficulty. For complex repositories, even the best generated trajectories may achieve relatively lower scores, causing a high threshold to discard many valuable hard instances. Conversely, lowering the threshold to preserve difficult cases may admit weak trajectories from simpler repositories, where near-perfect performance should be expected. This motivates a difficulty-aware filtering strategy that evaluates trajectory quality relative to the intrinsic difficulty of each repository.

To address this issue, we propose a difficulty-aware trajectory filtering strategy that sets instance-specific score thresholds according to the estimated complexity of each task. Specifically, we first introduce a difficulty estimation framework that assigns a fine-grained difficulty score to each document-to-repository generation instance. This score is then used to calibrate the filtering threshold, allowing the pipeline to retain valuable trajectories from challenging repositories while still enforcing stricter quality requirements for easier instances.

\subsection{Difficulty Scoring Framework}

\paragraph{Setup.} Let $\mathcal{I}$ denote the full set of benchmark instances. For each instance $i \in \mathcal{I}$, we capture three distinct difficulty signals:
\begin{itemize}[leftmargin=*, noitemsep, topsep=0pt]
    \item \textbf{Structural Signal} ($e_i \in \mathbb{Z}_{\geq 0}$): The total number of executable Python lines that fall within the scope of the target task.
    \item \textbf{LLM Signals} ($\ell_i^{(g)}, \ell_i^{(q)} \in \{1,2,3,4,5\}$): Independent 5-level qualitative difficulty judgments provided by two distinct LLM annotators ($g$ and $q$), both conditioned on the same task documentation.
\end{itemize}

For a subset $\mathcal{I}^{\star} \subseteq \mathcal{I}$ where agent rollouts have been executed, we additionally observe an empirical target signal $s_i \in [0,1]$, defined as the mean pass rate across all rollouts $\mathcal{R}_i$ collected for instance $i$:
\begin{equation}
s_i = \frac{1}{|\mathcal{R}_i|}\sum_{r \in \mathcal{R}_i} \text{score}(r)
\end{equation}
Intuitively, harder instances naturally yield lower rollout pass rates ($s_i$). Therefore, a well-formed difficulty estimator is expected to correlate negatively with $s_i$.

\paragraph{Component Score Normalization.} To eliminate scale disparities across heterogeneous signals, we map each raw indicator onto the unit interval $[0, 1]$:
\begin{itemize}[leftmargin=*]
    \item \textbf{Structural Score Optimization:} We apply a $\log(1+\cdot)$ transformation to the line counts $e_i$, followed by min-max normalization against the empirical $5^{\text{th}}$ and $95^{\text{th}}$ percentiles computed over the full pool $\mathcal{I}$:
    \begin{equation}
    d_i^{\text{struct}} = \operatorname{clip}_{[0,1]}\left( \frac{\log(1+e_i) - q_{0.05}}{q_{0.95} - q_{0.05}} \right)
    \end{equation}
    where $q_{\alpha} = \operatorname{Quantile}_{\alpha}\bigl(\{\log(1+e_j)\}_{j \in \mathcal{I}}\bigr)$.
    \item \textbf{LLM Score Alignment:} The ordinal levels from the annotators are linearly mapped onto the unit interval via:
    \begin{equation}
    \tilde{\ell}_i^{(m)} = \frac{\ell_i^{(m)} - 1}{4} \in \{0, 0.25, 0.5, 0.75, 1\}, \quad \forall m \in \{g, q\}
    \end{equation}
\end{itemize}

\paragraph{Multi-Feature Signal Fusion.} The final unified difficulty score $d_i$ is formulated as a convex combination of the three normalized components:
\begin{equation}
d_i = w_{\text{s}} \cdot d_i^{\text{struct}} + w_g \cdot \tilde{\ell}_i^{(g)} + w_q \cdot \tilde{\ell}_i^{(q)}
\end{equation}
\begin{equation*}
\text{s.t.} \quad w_{\text{s}}, w_g, w_q \geq 0, \quad w_{\text{s}} + w_g + w_q = 1
\end{equation*}
The convexity constraint guarantees that $d_i \in [0,1]$ while bounding each component to a mathematically comparable scale.

\paragraph{Weight Optimization.} The optimal weight vector $\mathbf{w}^* = (w_{\text{s}}^*, w_g^*, w_q^*)$ is calibrated on the sub-pool $\mathcal{I}^{\star}$ where both rollout target scores and LLM annotations are simultaneously available. Let $\mathbf{d}(\mathbf{w}) = \bigl(d_i(\mathbf{w})\bigr)_{i \in \mathcal{I}^{\star}}$ and $\mathbf{s} = (s_i)_{i \in \mathcal{I}^{\star}}$. We solve for the weights by maximizing the absolute Pearson correlation coefficient over the 2-simplex $\Delta^{2}$:
\begin{equation}
\mathbf{w}^* = \arg\max_{\mathbf{w} \in \Delta^{2}} \bigl| \rho\bigl(\mathbf{d}(\mathbf{w}),\, \mathbf{s}\bigr) \bigr|
\end{equation}
where $\Delta^{2} = \{\mathbf{w}: w_{\text{s}}+w_g+w_q = 1, \; w_\bullet \geq 0\}$. Since the objective function depends on $\mathbf{w}$ solely through this linear combination, the maximizer of $|\rho|$ is invariant to positive rescaling and bias. We compute the global optimum via an exhaustive 2D grid sweep over $(w_g, w_q)$ on a uniform dense mesh, setting $w_{\text{s}} = 1 - w_g - w_q$ and discarding out-of-bounds coordinates.

\subsection{Filtering Thresholds}

\begin{figure*}[t!] 
  \begin{center}
    \centerline{\includegraphics[width=1\textwidth]{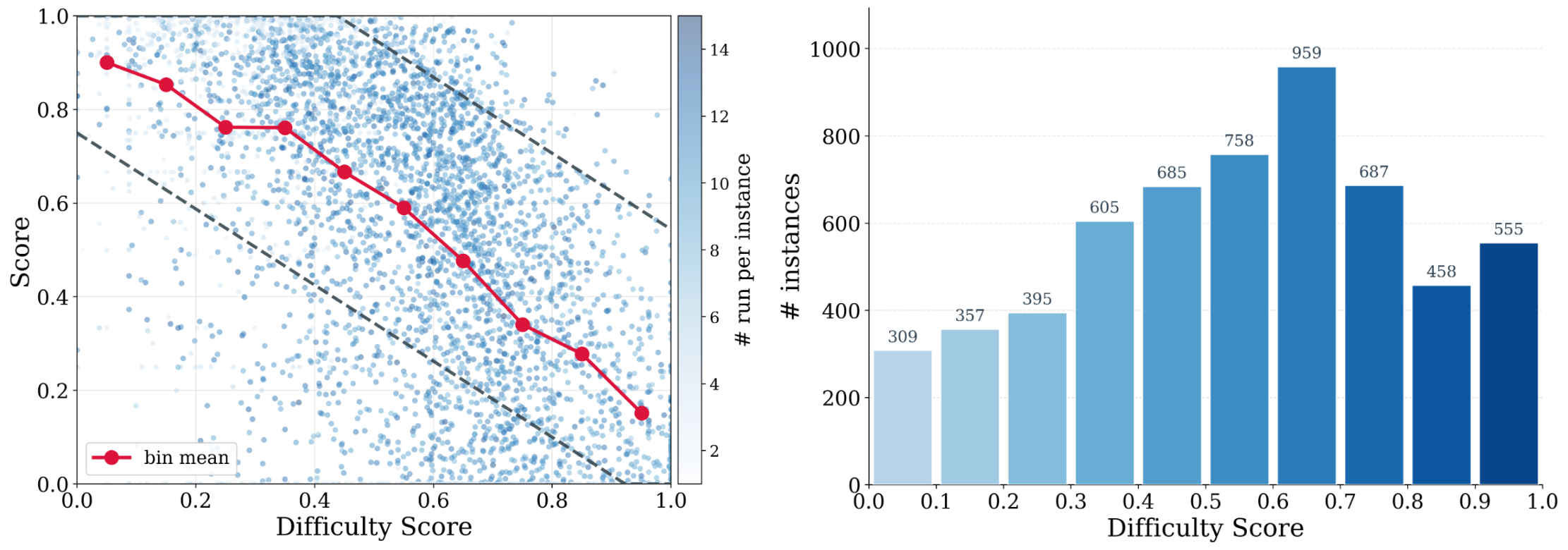}} 
    \caption{
Left: trajectory scores decrease as the estimated difficulty score increases, showing that fixed score thresholds would disproportionately discard trajectories from harder repository-generation instances. The red curve reports the mean trajectory score within each difficulty bin. Right: DeNovoSWE covers a broad range of task difficulties, with instances distributed across the full difficulty spectrum. These observations motivate our difficulty-aware trajectory filtering strategy, which adapts the filtering threshold according to instance difficulty.
    }
    \label{fig:difficulty_score}
  \end{center}
  \vspace{-0.6cm} 
\end{figure*}

Figure~\ref{fig:difficulty_score} illustrates the relationship between instance difficulty and trajectory score for trajectories distilled from DeepSeek-V4-Pro, together with the distribution of difficulty scores of DeNovoSWE. To implement difficulty-aware filtering, we partition the continuous difficulty range into five uniform intervals with a width of 0.2. We then assign an instance-specific trajectory score threshold to each difficulty interval. Easier instances are subject to stricter thresholds, since high pass ratios are expected for these tasks, whereas harder instances are assigned lower thresholds to retain informative trajectories that still solve a substantial portion of the repository. This dynamic thresholding scheme mitigates the selection bias introduced by a fixed global threshold, balancing trajectory quality with coverage over difficult repository-generation tasks. The mapping between difficulty intervals and their corresponding filtering thresholds is provided in Table~\ref{tab:score_threshold}.

As demonstrated in Table~\ref{tab:score_threshold}, the filtering threshold decreases monotonically as instance difficulty increases. For easier instances in the range of $[0.0, 0.2)$, we impose a high threshold of 0.90 to ensure strict quality control and filter out noisy or suboptimal rollouts. In contrast, for highly complex repositories in the $[0.8, 1.0]$ range, where fully successful execution is rarely achieved, the threshold is relaxed to 0.60. This relaxation enables the pipeline to retain informative long-horizon partial successes that still capture substantial repository-level reasoning and implementation behavior. Overall, this adaptive stratification balances execution fidelity with structural diversity, allowing DeNovoSWE to preserve high-quality trajectories while maintaining coverage over challenging repository-generation tasks.

\begin{table}[t]
    \centering
    \caption{Trajectory filtering thresholds for DeNovoSWE based on instance difficulty scores. A higher difficulty score indicates greater instance complexity.}
    \label{tab:score_threshold}
    \vspace{5pt}
    \begin{tabular}{cc}
        \toprule
        \textbf{Difficulty Score Range} & \textbf{DeNovoSWE Score Threshold} \\
        \midrule
        $[0.0, 0.2)$ & 0.90 \\
        $[0.2, 0.4)$ & 0.85 \\
        $[0.4, 0.6)$ & 0.80 \\
        $[0.6, 0.8)$ & 0.70 \\
        $[0.8, 1.0]$ & 0.60 \\
        \bottomrule
    \end{tabular}
\end{table}

\section{Experiments}
\subsection{Experiment Setup}

\textbf{Agent Scaffolding.} We employed OpenHands~\citep{wang2025openhands}, an open-source, event-driven platform, as the unified agent framework for all experiments. OpenHands facilitates LLM agents to iteratively edit files and execute shell commands within sandboxed containers. We selected this framework due to its proven ability to establish robust and reproducible baselines on benchmarks. We use AweAgent~\citep{aweagent2026} as the base to build all workflows.

\textbf{Trajectory data generation.} We employ DeepSeek-V4-Pro High~\citep{deepseekai2026deepseekv4} to generate execution trajectories for Supervised Fine-Tuning (SFT). First, we generate three independent trajectories for each instance. We then isolate the instances that fail to achieve a perfect score of 1.0 and perform another three inference rollouts for them. Finally, we filter the accumulated trajectories using our difficulty-aware filtering strategy to curate the final training dataset. Furthermore, during fine-tuning, we apply loss masking to the assistant's responses that correspond to failed tool invocations and heredoc operations. Ultimately, our filtering pipeline yields a final training set comprising approximately 11k high-quality trajectories. 

\textbf{Agent Post-training.} We perform post-training on the Qwen3-30B-A3B-Instruct~\citep{qwen3technicalreport} and Qwen3.5-35B-A3B~\citep{qwen3.5} as base model. The training process is configured with a learning rate of 1e-5, a batch size of 128, and a warmup ratio of 0.05, supporting a maximum context length of 131,072. 

\textbf{Evaluation Benchmarks and Metrics}
We conduct our evaluation on BeyondSWE~\citep{chen2026beyondswe} and NL2Repo-Bench~\citep{ding2025nl2repo}, two benchmarks specifically designed to assess the capability of generating entire repositories from scratch. For NL2Repo-Bench, all models are evaluated within the designated golden environment with all required dependencies pre-installed. To ensure statistical stability and mitigate experimental variance, all reported metrics are averaged across three independent execution trials. The detailed hyperparameter configurations employed during the evaluation phase are summarized in Table~\ref{tab:eval_param}.

\subsection{Experiment Results}
\begin{table*}[h]
    \centering
    \caption{Performance comparison on the BeyondSWE-Doc2Repo and NL2Repo benchmarks. For NL2Repo, all models are consistently evaluated within a consistent ground-truth (golden) environment for both agent execution and evaluation. All experiments were performed in triplicate, and the mean values are reported.}
    \label{tab:swe_bench_results}
    \vspace{10pt}
    \begin{tabular}{lcc}
        \toprule
        \textbf{Models} & \textbf{Doc2Repo} & \textbf{NL2Repo} \\
        \midrule
        
        \rowcolor{lightblue}
        \multicolumn{3}{c}{\textit{Proprietary Models}} \\
        
        GPT-5.4(CodeX)~\citep{openai2026gpt54} & 0.617 & - \\
        GPT-5.4~\citep{openai2026gpt54} & 0.563 & - \\
        DeepSeek-V4-Pro~\citep{deepseekai2026deepseekv4} & 0.566 & - \\
        GLM-5~\citep{glm5team2026glm5vibecodingagentic} & 0.568 & - \\
        Seed-Coder-2.0~\citep{seed2025seed} & 0.568 & - \\
        Qwen3.5-Plus~\citep{qwen3.5} & 0.524 & - \\
        Gemini3-Pro~\citep{googledeepmind2026gemini3pro}  & 0.520 & - \\

        \midrule
        \rowcolor{lightblue}
        \multicolumn{3}{c}{\textit{Qwen3-30B-A3B}} \\
        
        Qwen3-30B-A3B-Instruct~\citep{qwen3technicalreport} & 0.058 & 0.043 \\
        Scale-SWE-Agent~\citep{zhao2026immersion} & 0.292 & 0.183 \\
        \textbf{DeNovoSWE-Agent-30A3B} & 0.472 & 0.230 \\
        \midrule
        \rowcolor{lightblue}
        \multicolumn{3}{c}{\textit{Qwen3.5-35B-A3B}} \\
        Qwen3.5-35B-A3B~\citep{qwen3.5}  & 0.438 & 0.235 \\
        \textbf{DeNovoSWE-Agent-35A3B} & \textbf{0.500}  & \textbf{0.271} \\
        \bottomrule
    \end{tabular}
\end{table*}
We evaluate \textbf{DeNovoSWE-Agent-30A3B} and \textbf{DeNovoSWE-Agent-35A3B} against a broad set of competitive baselines on BeyondSWE-Doc2Repo and NL2RepoBench.

\textbf{Bridging the gap to proprietary frontier models.} Table~\ref{tab:swe_bench_results} shows that training on DeNovoSWE substantially improves open-weight agents for repository-scale generation. In particular, \textbf{DeNovoSWE-Agent-35A3B} achieves 50.0\% on BeyondSWE-Doc2Repo, narrowing the gap between open-weight models and strong proprietary baselines. Its performance is within 2.0 percentage points of Gemini3-Pro and 2.4 percentage points of Qwen3.5-Plus, despite using a substantially smaller open-weight backbone. This result demonstrates that high-quality long-horizon training data can substantially improve agents' ability to generate complete repositories from scratch.

\textbf{Substantial gains over open-weight baselines.} On the Qwen3-30B-A3B backbone, the original \texttt{Qwen3-30B-A3B-Instruct} model performs poorly on repository-generation tasks, achieving only 5.8\% on BeyondSWE-Doc2Repo and 4.3\% on NL2RepoBench. Training on issue-level SWE data with Scale-SWE-Agent~\citep{zhao2026immersion} improves the scores to 29.2\% and 18.3\%, respectively, showing that conventional SWE data scaling provides useful but limited transfer to whole-repository generation. In contrast, \textbf{DeNovoSWE-Agent-30A3B} further improves performance to 47.2\% on BeyondSWE-Doc2Repo and 23.0\% on NL2RepoBench. These gains demonstrate that training on DeNovoSWE substantially enhances long-horizon SWE capabilities for generating complete repositories from documentation.

We observe similar trends on the stronger Qwen3.5-35B-A3B backbone. The original \texttt{Qwen3.5-35B-A3B} already provides a strong baseline, achieving 43.8\% on BeyondSWE-Doc2Repo and 23.5\% on NL2RepoBench. After fine-tuning on DeNovoSWE, \textbf{DeNovoSWE-Agent-35A3B} further improves performance to 50.0\% and 27.1\%, respectively. These consistent gains across both Qwen3 and Qwen3.5 backbones indicate that DeNovoSWE provides effective long-horizon training data for whole-repository generation, rather than benefiting only a single model architecture.

\subsection{Ablation Study}
To evaluate the effectiveness of the proposed difficulty-aware trajectory filtering strategy, we conduct an ablation study against fixed, difficulty-independent score thresholds. Table~\ref{tab:threshold_ablation} reports the results on BeyondSWE-Doc2Repo and NL2RepoBench. This comparison allows us to examine whether adapting the filtering threshold to instance difficulty yields better training data than applying a single global threshold across all repository-generation tasks.

\begin{table}[h]
    \centering
    \caption{Evaluation results of downstream tasks across different trajectory filtering threshold strategies. The intervals represent the difficulty score ranges, while the corresponding inner values denote the filtering thresholds yielded by the model on the DeNovoSWE dataset. All reported metrics are averaged across three independent execution trials to ensure statistical stability.}
    \label{tab:threshold_ablation}
    \vspace{5pt}
    \begin{tabular}{ccccc@{\hspace{18pt}}cc} 
        \toprule
        \multicolumn{5}{c}{\textbf{Score Threshold per Difficulty Range}} & \textbf{Doc2Repo} & \textbf{NL2Repo} \\
        $\mathbf{[0.0, 0.2)}$ & $\mathbf{[0.2, 0.4)}$ & $\mathbf{[0.4, 0.6)}$ & $\mathbf{[0.6, 0.8)}$ & $\mathbf{[0.8, 1.0]}$ &  & \\
        \midrule
        \rowcolor{lightblue}
        \multicolumn{7}{c}{\textit{Difficulty Score Independent Filtering}} \\
        0.60 & 0.60 & 0.60 & 0.60 & 0.60 & 0.488 & 0.264 \\
        0.80 & 0.80 & 0.80 & 0.80 & 0.80 & 0.485 & 0.254 \\
        0.95 & 0.95 & 0.95 & 0.95 & 0.95 & 0.481 & 0.250 \\
        \midrule
        \rowcolor{lightblue}
        \multicolumn{7}{c}{\textit{Difficulty Score Dependent Filtering}} \\
        0.80 & 0.70 & 0.60 & 0.55 & 0.50 & 0.486 & 0.260 \\
        0.90 & 0.85 & 0.80 & 0.70 & 0.60 & \textbf{0.500} & \textbf{0.271} \\
        \bottomrule
    \end{tabular}
\end{table}

\textbf{Importance of dataset diversity.} 
The difficulty-independent baselines in Table~\ref{tab:threshold_ablation} apply a single trajectory-score threshold to all instances regardless of their difficulty. We observe that lowering this fixed threshold from 0.95 to 0.60 leads to consistent gains on both BeyondSWE-Doc2Repo and NL2RepoBench. This suggests that overly strict global filtering can remove useful trajectories from challenging repository-generation tasks. In this setting, retaining a broader range of hard instances can be more beneficial than selecting only high-scoring trajectories from easier repositories. This observation is consistent with Figure~\ref{fig:difficulty_score}, where trajectory scores tend to decrease as instance difficulty increases. A high global threshold therefore disproportionately filters out difficult repositories and narrows the training distribution. These results highlight the importance of preserving dataset diversity, especially coverage over challenging long-horizon instances, when constructing training data for whole-repository generation.

\textbf{Difficulty-aware filtering excludes low-quality trajectories.} The bottom half of Table~\ref{tab:threshold_ablation} reports the results of difficulty-aware filtering. Compared with the strongest difficulty-independent baseline, our difficulty-aware strategy improves performance from 0.488 to 0.500 on BeyondSWE-Doc2Repo and from 0.264 to 0.271 on NL2RepoBench. This suggests that difficulty-aware filtering not only preserves valuable trajectories from harder repositories, but also applies stricter quality control to easier instances. For low-difficulty tasks in the $[0.0, 0.2)$ range, a low trajectory score is less likely to reflect intrinsic task difficulty and more likely to indicate execution noise or an under-specified rollout. Consistent with this interpretation, the more permissive difficulty-aware configuration, which relaxes the easiest-tier threshold from 0.90 to 0.80, leads to lower downstream performance. This indicates that overly relaxed filtering can introduce low-quality trajectories from easy tasks, highlighting the need for stricter thresholds on low-difficulty instances while retaining more flexible criteria for challenging repository-generation tasks.

\section{Conclusion} 
In this work, we introduced \textbf{DeNovoSWE}, a large-scale real-world dataset for document-to-repository software engineering tasks, designed to support long-horizon code-agent training. Through a structured divide-and-conquer pipeline and an iterative critic-repair mechanism, DeNovoSWE automatically constructs 4,818 high-quality repository-generation instances, addressing the lack of verifiable training data for whole-repository generation. We further generated high-quality trajectories with DeepSeek-V4 and introduced a difficulty-aware trajectory filtering strategy to balance execution quality and task diversity.
Our empirical results demonstrate the effectiveness of DeNovoSWE for improving long-horizon SWE capabilities. Fine-tuning Qwen3-30B-A3B-Instruct on DeNovoSWE improves performance from 5.8\% to 47.2\% on BeyondSWE-Doc2Repo, surpassing the larger vanilla Qwen3.5-35B-A3B baseline. Experiments on the Qwen3.5-35B-A3B backbone show consistent gains, improving performance from 43.8\% to 50.0\% on BeyondSWE-Doc2Repo and from 23.5\% to 27.1\% on NL2RepoBench. These results show that high-quality long-horizon training data can substantially improve agents' ability to generate complete repositories from scratch. We believe DeNovoSWE provides a valuable resource for scalable repository-level task construction and for developing more capable long-horizon software engineering agents.

\bibliography{main}
\appendix
\newpage




\section{DeNovoSWE data structure}
\begin{table}[h]
    \centering
    \caption{DeNovoSWE data structure specification.}
    \label{tab:data_structure}
    \vspace{5pt}
    \begin{tabular}{ll}
        \toprule
        \textbf{Field} & \textbf{Description} \\
        \midrule
        instance\_id & Unique identifier for each benchmark instance. \\
        document & Ground-truth documentation provided to the agent for repository reconstruction. \\
        pypi\_name & PyPI package name used to enforce anti-cheating constraints during evaluation. \\
        image\_url & URL of the Docker image configured for the environment. \\
        user & GitHub username or organization owning the repository. \\
        repo & Name of the target GitHub repository. \\
        workdir & Working directory path of the repository inside the Docker container. \\
        unit\_test & List of all unit test identifiers. \\
        test\_patch & Code patch for unit tests, applied during the evaluation phase. \\
        test\_binary\_files & Binary files used by unit tests that are unsuitable for standard text patching. \\
        Difficulty & Difficulty score of the repository. \\
        \bottomrule
    \end{tabular}
\end{table}

\section{Hyperparameter}
The hyperparameters for SFT are detailed in Table~\ref{tab:sft_param}.

\begin{table}[h]
    \centering
    \caption{Key hyperparameters in the SFT phase.}
    \label{tab:sft_param}
    \vspace{5pt}
    \begin{tabular}{lr}
        \toprule
        \textbf{Parameter Name} & \textbf{Value} \\
        \midrule
        \texttt{Batch size} & 128 \\
        \texttt{Maximum Context Length} & 131,072 \\
        \texttt{Warmup ratio} & 0.05 \\
        \texttt{LR scheduler type} & Cosine \\
        \bottomrule
    \end{tabular}
\end{table}

To ensure reproducible and standardized evaluations across both the BeyondSWE-Doc2Repo and NL2Repo-Bench benchmarks, we establish a uniform set of core hyperparameters for all models' evaluation. These parameters are outlined in Table~\ref{tab:eval_param}. 

\begin{table}[h]
    \centering
    \caption{Hyperparameter configurations for agent evaluation on NL2Repo-Bench and BeyondSWE-Doc2Repo.}
    \label{tab:eval_param}
    \vspace{5pt}
    \begin{tabular}{lr}
        \toprule
        \textbf{Parameter Name} & \textbf{Value} \\
        \midrule
        \texttt{temperature} & 1.0 \\
        \texttt{max\_step} & 400 \\
        \texttt{max\_new\_token} & 32,768 \\
        \texttt{max\_token\_length} & 262,144 \\
        \bottomrule
    \end{tabular}
\end{table}

\section{License Filtering.}
During repository selection, DeNovoSWE applies license-aware filtering to exclude repositories whose licenses are unknown, missing, restrictive, or unsuitable for training-data construction. We retain only repositories under permissive licenses, such as MIT, Apache-2.0, BSD-family licenses, ISC, 0BSD, Unlicense, CC0-1.0, Zlib, PostgreSQL, NCSA, Boost-1.0, BSL-1.0, and Python-2.0. This ensures that the resulting dataset is constructed from open-source repositories with licenses more appropriate for research-oriented model training.

\section{Prompt for DeNovoSWE data construction.}\label{exp_detail}

\begin{markdownbox}[title={Repository overview prompt}]
You are analyzing repository `{{repo_name}}` in `{{workspace_dir}}`.

Goal:
- Write the opening overview section for the repository documentation.
- The final output must be natural documentation prose, not a checklist or capability inventory.
- Start from the provided README, repository tree, and capability outline, but inspect the codebase whenever the repository's role or architecture is still unclear.
- Prefer `str_replace_editor` with `command="view"` for concrete files and directories. Use `execute_bash` for search-oriented commands like `rg`, `find`, and `ls`.

Hard constraints:
- {search_guidance}
- Do not submit free-form prose in the chat. Finish with the structured `finish` tool payload.
- Do not write the overview into repository files.
- Do not merely restate the capability bullets; synthesize them into a coherent introduction.
- Do not rely only on traced symbols or README copy when the source tree suggests broader subsystems.

Overview writing rules:
- The section must begin with `## 1. Overview`.
- Lead with one or two short paragraphs explaining what the repository is, what problem domain it serves, and what kind of interface it exposes (for example: library, CLI, service, framework, plugin host, data pipeline).
- After the opening summary, include a short list of major workflows or subsystems when it improves clarity.
- Keep the focus on documentation-worthy repository behavior, not helper internals.
- Mention major interfaces such as CLI, Python API, server endpoints, configuration systems, or plugin surfaces only when supported by evidence.
- Avoid speculative claims, empty adjectives, and overly narrow function-level phrasing.
- Target roughly 120-350 words in total.

Suggested exploration flow:
1. Read the README and repository tree carefully.
2. Review the capability outline and identify the highest-level concepts it suggests.
3. Inspect package metadata and entry points such as `pyproject.toml`, `setup.py`, `package.json`, main `__init__` files, CLI modules, or service entry modules.
4. Open the main source packages under the likely source roots and confirm the repository's public workflows and core abstractions.
5. Draft an overview that feels like the beginning of real project documentation.

Few-shot example:
```markdown
## 1. Overview
The `target_repo` library is a comprehensive toolkit for evaluating the quality of biomolecular structure prediction models. It supports proteins, nucleic acids, and small molecules (ligands). It provides a unified interface to compare a "model" structure (prediction) against a "reference" structure (ground truth).

The core functionality includes:
- **Entity & Chain Mapping**: Automatically aligning entities (sequences) and chains between model and reference, handling permutations and symmetry.
- **Symmetry Resolution**: Resolving atom-level symmetry for residues and branched ligands to ensure optimal matching.
- **Multi-Metric Evaluation**: Computing LDDT, DockQ, pocket-aligned RMSD, clash scores, and structure validity checks.
- **Input Generation**: Converting between multiple structure and pipeline formats to support integration workflows.

The library exposes both a Command Line Interface (CLI) and a Python API.
```

Repository evidence:
- Instance ID: {{instance_id}}
- GitHub URL: {{github_url}}
- Likely source roots: {{likely_source_roots}}
- Capability count: {{capability_count}}

Capability outline:
{{capabilities_outline}}

README content:
```text
{{readme_content}}
```

Repository structure:
```text
{{repo_structure}}
```

Retry feedback:
{{retry_feedback}}

Before finishing:
1. Check whether the overview explains the repository's overall purpose rather than just enumerating modules.
2. Verify that any mentioned interfaces, workflows, or domains are supported by repository evidence.
3. Make sure the section reads like documentation opening prose, not analysis notes.
4. {pre_finish_guidance}
5. Submit the final overview section with the `finish` tool.
\end{markdownbox}

\begin{markdownbox}[title={Repository ability prompt.}]
You are a documentation architect for software repositories.
Your job is to derive an ordered repository capability outline from partial
evidence: README content, repository structure, and unit-test traced
symbols. The output will be used to plan structured documentation.
Return strict JSON only. No markdown fences. No extra prose.
JSON schema:
{
  "capabilities": [
    ["short capability phrase", "clear 1-2 sentence description"],
    ...
  ]
}
Example input:
Instance ID: acme_kvstore_demo
README:
```text
AcmeKV is a Python client for a remote key-value service. It supports
authenticated client setup, CRUD operations, namespace-based organization,
bulk writes, and exporting snapshots for backup or replication.
```
Repo structure:
```text
acme-kv/
  acmekv/
    client.py
    auth.py
    store.py
    namespace.py
    batch.py
    snapshot.py
  docs/
    index.md
  tests/
    test_store.py
```
Unique tested functions (deduplicated):
["acmekv.client.connect", "acmekv.auth.login", "acmekv.store.get",
 "acmekv.store.put", "acmekv.namespace.list_keys",
 "acmekv.batch.commit", "acmekv.snapshot.export"]
Unique tested classes (deduplicated):
["acmekv.client.Client", "acmekv.batch.BatchWriter"]
Example output:
{
  "capabilities": [
    [
      "Client authentication and connection setup",
      "Initialize the client, authenticate requests, and establish reusable sessions for interacting with the remote key-value service."
    ],
    [
      "Key-value read and write operations",
      "Create, update, fetch, and delete stored values through the core storage APIs that back routine data access."
    ],
    [
      "Namespace-based data organization",
      "Group and browse keys within namespaces so applications can separate logical datasets and manage them independently."
    ],
    [
      "Batch update workflows",
      "Stage and commit multiple write operations together to support bulk ingestion or coordinated updates."
    ],
    [
      "Snapshot export and backup flows",
      "Export repository state for backup, replication, or offline inspection using the snapshot-related interfaces."
    ]
  ]
}
\end{markdownbox}

\begin{markdownbox}[title={Ability document prompt}]
You are analyzing repository `{{repo_name}}` in `{{workspace_dir}}`.

Goal:
- Write one documentation section for exactly one ability: `{{ability_name}}`.
- Treat this run as a single-ability task. Do not draft or summarize the other abilities.
- Treat an ability as a coherent feature or workflow implemented by a set of related APIs. The final section should read like realistic repository documentation, not like disconnected micro-docs for random helpers.
- Start from the README content and repository structure shown below, then inspect the local code and related unit tests to verify signatures and behavior before finishing.
- Prefer `str_replace_editor` with `command="view"` for concrete files and directories. Use `execute_bash` for search-oriented commands like `rg`, `find`, `ls`, and `sed`.

README content:
```text
{{readme_content}}
```

Repository structure:
```text
{{repo_structure}}
```

Hard constraints:
- {search_guidance}
- The README content and repo structure above are mandatory grounding context for this run.
- Do not submit free-form prose in the chat. Finish with the structured `finish` tool payload.
- Do not write the documentation into repository files.
- Every qualified symbol name listed under `Current ability symbol bundle` must appear verbatim in the final section.
- Every name listed under `Highlighted exact APIs` must receive an exact, code-grounded API description. If defaults, constants, or required parameters materially affect behavior, mention them.
- Inspect unit tests related to the current ability whenever they exist. Do not make downstream models guess things that the tests rely on, including exact signatures, raised errors, return-field names, required constants, mode strings, or configuration keys.
- If the tests or public call sites depend on exact behavior such as a specific `ValueError`, payload key, or default branch, document that contract explicitly.
- Check the code for exact parameter names, method names, and class ownership. Never guess or rename an API.
- Keep the section documentation-like, not a line-by-line implementation dump. Avoid spilling low-level logic unless it is needed to explain a tested or user-visible contract.
- If a detail is not supported by repository code, tests, or clearly linked official docs, do not assert it.

Writing rules:
- The section must start with `## {{ability_name}}`.
- Open with one short paragraph that explains the current ability at workflow/component level.
- After the overview, organize the section into one or more component-level blocks such as `### Component: ...`.
- Inside each component block, describe the important classes and functions that jointly implement that part of the ability. Use the exact import path and exact API names.
- Mention all assigned symbols, but spend most detail on the highlighted APIs and absorbed direct functions because their exact API surface is evaluation-critical.
- Use subsections like `Import Path`, `Description`, `Signature`, `Parameters`, `Returns`, `Raises`, `Methods`, and `Notes (Logic & Behavior)` when they help. Do not force every subsection for every API.
- For classes, document the class role first, then focus on the methods that matter to the ability. For standalone functions, document parameters, return shape, and important errors clearly.
- Include constants, modes, or defaults only when they are part of the effective contract or are hard to infer correctly without documentation.
- Avoid heavy overlap with earlier completed ability sections.
- You may use short fenced Python signatures when that is the clearest way to pin down an evaluation-critical API. Do not include long code excerpts or implementation snippets.
- Target roughly 250-900 words.

Formatting guidance:
- Think in this order: ability overview -> component -> class/function -> method-level details only where needed.
- The goal is to make the repo reconstructable without turning the task into implementation copy-paste.
- Prioritize exactness for APIs that are directly exercised by tests. Do not waste space exhaustively documenting trivial helpers.

Few-shot formatting examples:
The examples below show good component-level structure for the body of the section after the top-level `## {{ability_name}}` heading. Use them as style references, not as a fixed schema.

Example 1:
```example
### Component: Credential Management

**Import Path:**
`from target_repo.credentials import VerifiableCredential`

**Description:**
Handles the lifecycle of Verifiable Credentials (VCs): building, signing, and verifying. It uses JSON-LD normalization and Ed25519 signatures.

#### Class: `VerifiableCredential`

**Signature:**
```python
class VerifiableCredential:
    def __init__(self, data: Optional[Dict[str, Any]] = None): ...
```

**Description:**
Main class for manipulating VCs. Can be initialized with a dictionary matching the VC schema.

**Methods:**

1. **`build(...)`**
   **Signature:**
   ```python
   def build(
       self,
       issuer_did: str,
       subject_did: str,
       credential_id: Optional[str] = None,
       types: Optional[List[str]] = None,
       contexts: Optional[Union[List, str]] = None,
       issuance_date: Optional[datetime] = None,
       expiration_date: Optional[datetime] = None,
       credential_subject: Optional[Dict[str, Any]] = None
   ) -> "VerifiableCredential": ...
   ```
   **Description:** Constructs the unsigned VC data structure. Populates fields like `issuer`, `issuanceDate`, and `credentialSubject`.

2. **`sign(...)`**
   **Signature:**
   ```python
   def sign(
       self,
       issuer_signing_key: nacl.signing.SigningKey,
       verification_method_id: str,
       proof_purpose: Optional[str] = "assertionMethod"
   ) -> "VerifiableCredential": ...
   ```
   **Description:** Signs the VC.
   **Logic:**
   1. Canonicalizes the document using **URDNA2015**.
   2. Signs the hash using the `issuer_signing_key` (Ed25519).
   3. Encodes the signature in **base58btc**.
   4. Appends a `proof` object (type `Ed25519Signature2020`) to the VC.

3. **`verify_signature(...)`**
   **Signature:**
   ```python
   def verify_signature(
       self,
       issuer_public_key: nacl.signing.VerifyKey,
       expected_issuer_did: Optional[str] = None,
       expected_subject_did: Optional[str] = None
   ) -> bool: ...
   ```
   **Description:** Verifies the cryptographic integrity of the VC.
   **Logic:**
   1. Removes the `proofValue` from the proof block.
   2. Canonicalizes the document (URDNA2015).
   3. Verifies the signature against the canonical doc using `issuer_public_key`.
   4. Optionally checks if `issuer` and `credentialSubject.id` match expected values.

4. **Serialization & Factories**
   `to_dict() -> Dict`: Exports the VC to a Python dictionary.
   `to_json() -> str`: Exports the VC to a JSON string.
   `@classmethod from_dict(data: Dict) -> "VerifiableCredential"`: Creates a VC instance from a dictionary.
   `@classmethod from_json(json_str: str) -> "VerifiableCredential"`: Creates a VC instance from a JSON string.

---
```

Example 2:
```example
### Component: `fetch_gerrit_change`

**Import Path:**
`from target_repo.server import fetch_gerrit_change`

**Signature:**
```python
def fetch_gerrit_change(
    ctx: Context,
    change_id: str,
    patchset_number: Optional[str] = None,
    include_comments: bool = False
) -> Dict[str, Any]: ...
```

**Description:**
Retrieves comprehensive details about a specific Gerrit change. This includes project metadata, revision history, modified files, and optionally inline comments.

**Parameters:**
* **ctx** (*Context*): The MCP request context (provided by the runtime).
* **change_id** (*str*): The unique identifier of the Gerrit change.
* **patchset_number** (*str, optional*): The specific patch set number to fetch. If `None`, defaults to the latest revision.
* **include_comments** (*bool, optional*): If `True`, fetches existing inline comments for the target revision. Default: `False`.

**Returns:**
* **result** (*Dict[str, Any]*): A dictionary containing:
    * `change_info`: Raw metadata from Gerrit.
    * `project`: The project name.
    * `revision`: The commit hash of the target revision.
    * `patchset`: Detailed info about the patch set.
    * `files`: A list of modified files (including diff stats).
    * `inline_comments`: (Optional) Existing comments on the code.
    * `excluded_large_files`: (Optional) List of files skipped due to exclusion patterns.

**Raises:**
* **ValueError**: If the change ID or patch set number is invalid or not found.

**Notes (Logic & Behavior):**
* **Data Retrieval**: Queries the Gerrit REST API endpoint `a/changes/{{{{change_id}}}}/detail` requesting options for current/all revisions, commits, messages, labels, and accounts.
* **Patch Set Resolution**: If `patchset_number` is provided, it iterates through available revisions to find the matching number. Otherwise, it uses `current_revision`.
* **File Filtering**: Checks against exclusion patterns (configured via `GERRIT_EXCLUDED_PATTERNS`). Files matching these patterns are skipped to prevent processing large or irrelevant files, such as `node_modules` or lock files.
* **Diff Fetching**: For each non-excluded file, it fetches the diff using `a/changes/.../diff`.
* **Error Handling**: Raises informative errors if the project or revision data is missing from the API response.

---
```

Repository evidence:
- Repo instance ID: {{repo_instance_id}}
- Current ability order: {{ability_index}} / {{ability_count}}
- GitHub URL: {{github_url}}
- Likely source roots: {{likely_source_roots}}

Full ability outline:
{{capabilities_outline}}

Current ability description:
{{ability_description}}

Previously completed ability docs:
{{previous_ability_docs}}

Current ability symbol bundle:
{{current_ability_symbols}}

Must mention these exact names:
{{must_mention_names}}

Highlighted exact APIs:
{{highlighted_api_names}}

Retry feedback:
{{retry_feedback}}

Before finishing:
1. Inspect the local source files for the highlighted APIs and any ambiguous symbol names.
2. Inspect related unit tests, call sites, and fixtures for the current ability when they exist.
3. Verify that exact names, signatures, defaults, ownership relations, raised errors, and return structures are correct.
4. Make sure every required symbol name appears in the section and the focus stays on the current ability only.
5. {pre_finish_guidance}
6. Submit the final section with the `finish` tool using the current ability name and the markdown body.
\end{markdownbox}

\end{document}